\documentclass[11pt]{article}
\usepackage{amssymb}
\newtheorem{pro}{Proposition}
\def\sG{{\scriptscriptstyle G}}
\def\bp{\noindent{\em Proof}.\ }
\def\ep{\hfill$\square$\vspace*{1mm}}
\title{
 Haldane--Wu statistics and Rogers dilogarithm
 }
\author{\normalsize\sc Andrei G.~Bytsko }
\date{\small November, 2002   \\ 
 \small math-ph/0211026 }
\begin{document}
\maketitle
%
\abstract{
The Haldane--Wu exclusion statistics is considered
{}from the generalized extensive statistics point 
of view and certain related mathematical aspects
are investigated. A series representation for the 
corresponding generating function is proven.
Equivalence of two formulae for the central charge,
derived for the Haldane-Wu statistics via
the thermodynamic Bethe ansatz, is established.
As a corollary, a series representation with
a free parameter for the Rogers dilogarithm is found.
It is shown that the generating function, the
entropy, and the central charge for the Gentile 
statistics majorize those for the Haldane--Wu 
statistics (under appropriate choice of parameters).
{}From this, some dilogarithm inequality is derived.
}
\section{Introduction}
Consider (1+1)-dimensional system of relativistic
particles on an interval of length~$L$. If the
particle interaction is described by a factorizable
scattering matrix then the boundary condition for 
the wave function of a particle has the form
\begin{equation}\label{BA}
 \exp (i L m_k \sinh \theta_k )\prod_{l\neq k}^{N} S_{kl}
 (\theta_k - \theta_l )= \varsigma_k \,, \qquad k=1,\ldots,N \,,
\end{equation}
where $\theta_k$ and $m_k$ are the rapidity and the mass of 
the particle, $S_{kl}(\theta)$ is the two--particle scattering 
matrix, and $N$ is the total number of particles. The phases 
$\varsigma_k$ can be different for different particles
(their exact values are not relevant for our purposes).
For simplicity we consider the case when all particles
belong to the same species and have mass~$m$.

Analysis of the multiparticle system (\ref{BA}) in
the thermodynamic limit ($L \rightarrow \infty$, but
the density $N/L$ remains finite) is based on the
thermodynamic Bethe ansatz~\cite{KM}.
Apart from the system (\ref{BA}) it uses the thermodynamic
equilibrium condition, i.e., the condition of minimum
of the free energy $\cal F$ 
(${\cal F}={\cal E} - T {\cal S}$, where $T$ --- the 
temperature, $\cal E$ --- the total energy, $\cal S$ --- 
the entropy of the system). Thus, the initial data
for the thermodynamic Bethe ansatz
are the two--particle scattering matrix $S(\theta)$, 
the spectrum of particle masses, and the statistics 
which governs filling in states in the momentum 
space. The latter, so--called {\em exclusion statistics},
determines the exact form of the entropy of the system.

For one-dimensional systems, the exclusion statistics
is not necessarily of fermion or boson type but can
depend nontrivially on the number of particles already
present in a given state. For instance, a generalized 
{\em extensive} statistics is defined by a choice of
generating function $f(t)$ such that
\begin{equation}\label{fgen}
 \bigl( f(t) \bigr)^N = \sum_{n \geq 0} W(N,n) \, t^n \,,
\end{equation}
where $W(N,n)$ -- the number of possible ways for $n$
identical particles to occupy $N$ states. It is natural 
to impose the condition $f(0)=1$ that implies that the 
vacuum is realized with the probability one independently
on the size of a system.

The thermodynamic Bethe ansatz allows one to obtain 
certain information about
the ultra-violet (i.e., higher temperature) limit of
the system under consideration. In particular, it allows
one to find the effective central charge for the 
corresponding conformal model. For instance, in the case
of a generalized extensive statistics, the effective 
central charge is given by the following formula \cite{By}
\begin{equation}\label{c2}
 c = \frac{6}{\pi^2} \Bigl[ 
 \int_0^{x_0} \frac{ dt }{t} \ln f(t) 
 - \frac{1}{2} \ln x_0 \ln f (x_0)\Bigr] \,.
\end{equation}
Here $x_0$ is the positive root of the equation
\begin{equation}\label{cG2}
  \ln x_0 + \Phi \, \ln f(x_0) =0 \,,
\end{equation}
which is unique if $f(t)$ is monotonically increasing
and $\Phi \geq 0$. From the physical point of view,
$\Phi$ is related to the asymptotics of the scattering
matrix, $2\pi i \Phi= \ln S(-\infty) - \ln S(\infty)$,
but we will treat $\Phi$ just as a free
non-negative parameter.

\section{Haldane--Wu statistics}
The Haldane--Wu statistics \cite{Ha,Wu} is one of the 
most studied cases of an exotic statistics (see, e.g.,
\cite{Wu,VO,Is,HW,Po,BF}). It has applications, 
for instance, in the quantum Hall effect theory. For
this statistics, the number of possible ways for $n$
identical particles to occupy $N$ states is given by  
\begin{equation}\label{W1}
  W_g(N,n) = 
   \frac{(N+ (1-g)n +g-1)!}{n! \, ( N - g n + g-1)!} \,,
\end{equation}
where $0\leq g \leq 1$. The Haldane--Wu statistics
interpolates between fermions ($g=1$) and bosons ($g=0$). 

The Haldane--Wu statistics is {\em asymptotically}
extensive in the following sense. For a generalized
extensive statistics (\ref{fgen}), the entropy density
is defined as 
\begin{equation}\label{ent1}
 s(\mu) =
 \lim_{N \rightarrow \infty} \frac{1}{N} \ln W(N,\mu N) \,.
\end{equation}
One can show that (see, e.g., \cite{By})
\begin{equation}\label{ent2}
 s(\mu) = \ln f(x) - \mu \, \ln x \,,
\end{equation}
where $x\equiv x(\mu)$ is the positive root of the equation
(the prime denotes a derivative)
\begin{equation}\label{xeq}
  x \, f^\prime (x) = \mu \, f(x) \,.
\end{equation}
It follows then that
\begin{equation}\label{le}
 f\bigl(x(\mu)\bigr) \equiv  f(\mu) = 
 \exp \{ s(\mu) - \mu \, \partial_\mu s(\mu) \} \,.
\end{equation}
In the case of the Haldane--Wu statistics, application
of the Stirling formula to (\ref{W1}) yields
\begin{equation}\label{ent3}
 s_g(\mu)  = (1+\mu(1-g))\ln (1+\mu(1-g)) -
 \mu\ln \mu - (1-g\mu) \ln (1-g\mu) \,.
\end{equation}
Now, comparison with (\ref{le}) shows that
\begin{equation}\label{fm}
 f_g(\mu) = \frac{1+(1-g)\mu}{1-g\mu} 
\end{equation}
and, therefore, equation (\ref{xeq}) acquires the form
\begin{equation}\label{diff2}
 \bigl( g f_g(t) + 1-g \bigr) \, t \, f_g^\prime(t) = 
 f_g^2(t) - f_g(t)  \,.
\end{equation}
Whence, determining the integration constant from the
condition $f_g(0)=1$, we obtain
\begin{equation}\label{ft}
  f_g(t) - 1 = t \, \bigl( f_g(t) \bigr)^{1-g} \,.
\end{equation}
If $f^{1-g}$ on the r.h.s.~is understood as 
$\exp[(1-g)\ln f]$, where $\Im (\ln f) =0$ for $f>0$, 
then for $0 \leq g \leq 1$ equation (\ref{ft}) has unique
positive solution. Equations (\ref{fm}) and (\ref{ft})
are well-known in the context of exotic exclusion 
statistics \cite{Wu,VO,Is}.

Notice that the solution to (\ref{ft}) satisfies
a duality relation:
\begin{equation}\label{fdual}
  f_g(t) \, f_{1-g}(-t) = 1 \,.
\end{equation}
Furthermore, it follows from (\ref{diff2}) that 
$ t f_g^\prime / (f_g-1) > 0$, that is $f_g(t)$ 
is a monotonically increasing function. 
{}From  (\ref{ft}) we infer (with the help of
the $g>1$ counterpart of (\ref{yi})) also that 
\begin{equation}\label{fas}
 f_g(t) < \frac 1g + t^{\frac 1g} 
\end{equation}
for non-negative $t$. Actually, the r.h.s.~of (\ref{fas})
gives the asymptotics of $f_g(t)$ for large~$t$.

Using equation (\ref{ft}), we can compute derivatives
of $f_g$ at $t=0$ in a recursive way:
\begin{equation}\label{rec}
 f_g^{(n)} (0) = n \, \partial_t^{n-1} 
   (f_g^{1-g})\vert_{t=0} \,.
\end{equation}
First few values allow us to conjecture that
$f_g$ is given by the following Taylor series
\begin{equation}\label{fn}
 f_g(t) = 1 + t + \sum_{n=2}^\infty 
 \Bigl( \prod_{k=2}^{n} \bigl( 1- \frac{gn}{k} \bigr) 
 \Bigr) \, {t^n} \,.
\end{equation}
This series for $f_g$ was suggested in \cite{Is};
some combinatorial arguments were given for it in
\cite{Po} (for positive integer values of $g$).
Furthermore, it was also suggested in \cite{Su,VO,Po}
that logarithm of~$f_g$ is given by the series 
\begin{equation}\label{wn}
 \ln f_g(t) = t + \sum_{n=2}^\infty 
 \Bigl( \frac 1n \prod_{k=1}^{n-1} 
 \bigl( 1- \frac{gn}{k} \bigr) \Bigr) \, {t^n} \,.
\end{equation}
We will prove the following statement.
\begin{pro}\label{P1}
The series (\ref{fn}) and (\ref{wn}) are absolutely
convergent for 
\begin{equation}\label{t0}
 \ln |t|< \ln t_0 = - g \ln g - (1-g) \ln (1-g) \,.
\end{equation}
On this interval, the series (\ref{fn}) and (\ref{wn})
are, respectively, the positive solution of 
equation (\ref{ft}) and its logarithm. Moreover,
for an integer~$m$ we have on the same interval  
\begin{equation}\label{fmn}
 \bigl( f_g(t) \bigr)^m = 1 + mt + \sum_{n=2}^\infty 
 \Bigl( m \prod_{k=2}^{n} 
 \bigl( 1 + \frac{m-1-gn}{k} \bigr) \Bigr) \, {t^n} \,.
\end{equation}
\end{pro}

\bp Let $f_n$ and $w_n$, $n=0,1,2,\ldots$ denote, 
respectively, the coefficients of $t^n$ in the
series (\ref{fn}) and (\ref{wn}) (so that $w_0=0$ and
$f_0=f_1=w_1=1$). Notice that they can be written in 
terms of the gamma--function:
\begin{eqnarray}\label{fw1}
 f_n &=& \frac{\Gamma(1+(1-g)n)}{n! \, \Gamma(2-gn)} 
 = -\frac{\sin \pi g n}{\pi \, n!} \, 
 \Gamma(1+(1-g)n)\, \Gamma(gn-1) \,, \\
 \label{fw2}
 w_n &=& \frac{\Gamma((1-g)n)}{n! \, \Gamma(1-gn)} 
 = \frac{\sin \pi g n}{\pi \, n!} \,
 \Gamma((1-g)n)\, \Gamma(gn) \,.
\end{eqnarray}
Let us denote
$\tilde{f}_n = f_n / \sin \pi g n $ and
$\tilde{w}_n = w_n /  \sin \pi g n $.
Applying the Stirling formula (for large $z$ and
$\delta \ll z$) in the form  
$\ln \Gamma(z+\delta) - \ln \Gamma(z) = 
 \delta \ln z + o(1)$,  we find
\begin{equation}\label{lim} 
 \lim_{n \rightarrow \infty} \ln \Bigl| 
    \frac{\tilde{f}_{n+1}}{\tilde{f}_n} \Bigr| =
 \lim_{n \rightarrow \infty} \ln \Bigl| 
    \frac{\tilde{w}_{n+1}}{\tilde{w}_n} \Bigr| =
   g \ln g + (1-g) \ln (1-g) \,.
\end{equation}
Thus, the series $\sum_{n\geq 1} \tilde{f}_n t^n$ 
and $\sum_{n\geq 1} \tilde{w}_n t^n$ and, hence, 
the series (\ref{fn})--(\ref{wn}) converge 
absolutely on the interval~(\ref{t0}).

In order to prove the second assertion of the
proposition we observe that equation (\ref{diff2}), 
being multiplied by $f_g^{m-2}$, acquires the form
\begin{eqnarray}
 \label{diff3b}
  m=1 : &&
  (1-g) t \, (\ln f_g)^\prime = f_g -1
     - g t \, f_g^\prime \,, \\
 \label{diff3a}
  m \neq 0,1 : &&
  f_g^m - \frac{g}{m} \, t \, \bigl(f_g^m\bigr)^\prime = 
  f_g^{m-1} + \frac{(1-g)}{m-1} \, t \, 
   \bigl(f_g^{m-1}\bigr)^\prime \,.
\end{eqnarray}
Similarly, for the function $h_g(t)=f_g(t)-1$ equation
(\ref{diff2}) yields
\begin{eqnarray}\label{diff3c}
  m \neq 0,1 : &&
  h_g^m - \frac{g}{m} \, t \, \bigl(h_g^m\bigr)^\prime = 
  - h_g^{m-1} + \frac{1}{m-1} \, t \, 
   \bigl(h_g^{m-1}\bigr)^\prime \,.
\end{eqnarray}
{}From (\ref{diff3b})--(\ref{diff3b}) we derive 
relations between the Taylor coefficients
\begin{eqnarray}
 \label{fw}
 && w_n = \frac{1-gn}{(1-g)n} \, f_n 
  \,, \qquad n=1,2,\ldots \,, \\
 \label{ff}
 && f^{[m]}_n = 
   \frac{m(m-1+(1-g)n)}{(m-1)(m-gn)} \, f_n^{[m-1]} 
   \,, \qquad n=0,1,\ldots \,, \\
 \label{hh}
 && h^{[m]}_n = 
   \frac{m(n+1-m)}{(m-1)(m-gn)} \, h_n^{[m-1]} 
   \,, \qquad n \geq m =2,3,\ldots \,.
\end{eqnarray}
Here $f^{[m]}_n$ and $h^{[m]}_n$ are, respectively, 
Taylor coefficients of the series 
$\bigl(f_g(t)\bigr)^m= \sum_{n\geq 0} f^{[m]}_n t^n$
and 
$\bigl(h_g(t)\bigr)^m= \sum_{n\geq m} h^{[m]}_n t^n$.
Solving equations (\ref{ff})--(\ref{hh}), we find
\begin{eqnarray}
 && \label{fff} 
 f^{[m]}_n = m \, f_n \, \frac{\Gamma(2-gn)\,\Gamma(m+(1-g)n)}%
 {\Gamma(1+(1-g)n)\,\Gamma(m+1-gn)} \,, \\
 && \label{hhh}
 h^{[m]}_n = m \, h^{[1]}_n \, \frac{(n-1)! \,\Gamma(2-gn)}%
 {(n-m)! \,\Gamma(m+1-gn)} \,.
\end{eqnarray}
Substituting $m=n$ into (\ref{hhh}) and taking
into account that $h^{[1]}_n=f_n$ and $h^{[n]}_n=1$
for all $n\geq 1$, we obtain exactly formula (\ref{fw1})
for the coefficients of the series (\ref{fn}).
The assertion that the series (\ref{wn}) is logarithm 
of the series (\ref{fn}) follows now {}from the 
relation~(\ref{fw}). Finally, combining (\ref{fff})
with (\ref{fw1}), we find the formula 
\begin{equation}\label{fffff} 
 f^{[m]}_n = \frac{m\,\Gamma(m+(1-g)n)}{n! \,
 \Gamma(m+1-gn)} \,, \qquad n=1,2,\ldots
\end{equation}
that yields the series expansion~(\ref{fmn}). 
Analysis of absolute convergence of this series on
interval (\ref{t0}) is done in the same way as for
series (\ref{fn}) and~(\ref{wn}). Although we have
considered only positive values of $m$, an easily 
verified relation $(-1)^n f^{[m]}_{1-g,n}=
 f^{[-m]}_{g,n}$ together with the duality 
relation (\ref{fdual}) show that (\ref{fmn}) holds
for negative $m$ as well.
\ep

Let us remark that, if we assume validity of (\ref{fff}) 
for $m=1-g$, then we can use relation 
$f_{n+1} = f^{[1-g]}_n$ (that follows from (\ref{ft}))
to obtain a recurrence relation. Solution of this
relation coincides with~(\ref{fw1}). This indicates that 
formula (\ref{fmn}) holds also for non-integer~$m$. 
Another evidence for this is that series
(\ref{wn}) and (\ref{fmn}) are consistent in the sense
that $\lim_{m\rightarrow 0} (f^m_g -1)/m = \ln f_g$. 

\section{Central charge for Haldane--Wu statistics}

Strictly speaking, formula (\ref{W1}) for counting of
states in the Haldane--Wu statistics needs additional
conventions for finite $n$ and~$N$. It however is
sufficient for constructing the corresponding
thermodynamic Bethe ansatz 
along the same lines as in the case of the ordinary
statistics. This approach does not use explicit 
form of $f_g$ and leads to the following expression
for the effective central charge \cite{BF}
\begin{equation}\label{cHW}
 c_g = \frac{6}{\pi^2}  L (y_0) \,,   
\end{equation}
where $y_0$ is the positive root of the equation
\begin{equation}\label{cTBA}
 \ln y_0 = ( \Phi + g ) \, \ln (1-y_0) \,.  
\end{equation}
The r.h.s.~of (\ref{cHW}) contains the Rogers 
dilogarithm that is defined as
\begin{equation}\label{dil}
  L(x) = - \frac 12  \int_{0}^{x}  dt \, \Bigl(
  \frac{\ln (1-t)}{t} + \frac{\ln t}{1-t} \Bigr) =
  \sum_{n=1}^{\infty} \frac{x^n}{n^2} +
   \frac 12  \ln x \ln (1-x) \,.
\end{equation}

On the other hand, since the Haldane--Wu statistics
is asymptotically extensive, the corresponding 
effective central charge $c_g$ should also be given
by the general formula (\ref{c2}) if we 
substitute~$f=f_g$. Thus, we have two expressions,
rather different at the first site, for the  
effective central charge in the Haldane--Wu statistics.
Since the thermodynamic Bethe ansatz
derivation of the formula for an effective central 
charge involves a nontrivial limit and uses some
additional assumptions, it appears to be instructive
to provide a direct proof of equivalence of the two
expressions for~$c_g$.
\begin{pro}\label{P2} 
Let $0 \leq g \leq 1$, $\Phi \geq 0$,
and $f_g(t)$ be the positive solution of 
equation~(\ref{ft}). Then the following equality holds
\begin{equation}\label{pr1} 
 \int_0^{x_0} \frac{ dt }{t} \ln f_g(t) 
  - \frac{1}{2} \ln x_0 \ln f_g (x_0) 
 = L \Bigl(1- \frac{1}{f_g(x_0)} \Bigr) = L(y_0) \,,
\end{equation}
where $y_0$ is the positive root of equation (\ref{cTBA}),
and $x_0$ is the positive root of equation
\begin{equation}\label{cTBA2}
  \ln x_0 + \Phi \, \ln f_g(x_0) =0 \,.
\end{equation}
\end{pro}

\bp 
Notice that, since $f_g(t)$ increases monotonically,
equation (\ref{cTBA2}) has unique positive 
solution~$x_0$. Furthermore, $x_0 \leq 1$ 
because $f_g(0)=1$.

Consider the function $y(t)=1-1/f_g(t)$. It allows
us to rewrite equation (\ref{ft}) as $t=y(1-y)^{-g}$. 
Therefore
\begin{eqnarray}
 \label{cv}
 && \int \frac{ dt }{t} \ln f_g(t) 
  - \frac{1}{2} \ln t \ln f_g (t) =
 - \int d \bigl(\ln y - g \ln(1-y) \bigr) \, \ln(1-y) \\
 \nonumber   && \qquad 
  + \frac{1}{2} \bigl(\ln y - g \ln(1-y) \bigr) \, \ln(1-y) =
  - \int \frac{ dy }{y} \ln (1-y) + 
  \frac{1}{2} \ln y \ln(1-y) \,.
\end{eqnarray}
Comparison of the last expression with the definition
(\ref{dil}) yields the first equality in~(\ref{pr1}). 
Further, employing equations (\ref{ft}) and (\ref{cTBA2}), 
we obtain
\begin{eqnarray}
 \nonumber
 \ln y(x_0) &=& \ln(f_g(x_0) -1) - \ln f_g(x_0) =
  \ln x_0 -g \ln f_g(x_0)  \\
 \label{z0}
  &=& - (\Phi+g) \ln f_g(x_0) = (\Phi+g) \ln (1 - y(x_0)) \,.
\end{eqnarray}
Since (\ref{cTBA}) has unique positive solution for
$(\Phi+g) \geq 0$, we conclude that $y(x_0)=y_0$, which 
proves the second equality in~(\ref{pr1}). \ep

Let us now formulate a mathematical corollary of
Propositions~\ref{P1} and~\ref{P2}.
\begin{pro}\label{Psum}
Let $0 < g < 1$ and $\Phi\geq 0$ and
 let $y_0$ be the positive root of~(\ref{cTBA}). Then
\begin{equation}\label{Gsum}
 \sum_{n=1}^\infty \sin \pi g n \,
 \frac{\Gamma((1-g)n)\, \Gamma(gn)}{\pi \, n \, n!} \, 
 \Bigl( y_0 (1-y_0)^{-g} \Bigr)^n  +  
 \frac{\Phi}{2} \bigl( \ln (1-y_0) \bigr)^2  = L(y_0)
\end{equation}
if $t=y_0 (1-y_0)^{-g}$ satisfies condition~(\ref{t0}).
\end{pro}

\bp
Indeed, by Proposition~\ref{P1} we can substitute
the series (\ref{fn}) into the integral on the 
l.h.s.~of (\ref{pr1}) and carry out term--wise 
integration. The resulting series converges to 
the value of the integral if the condition of
absolute convergence (\ref{t0}) is satisfied.
The quantity $x_0$ entering the l.h.s.~of (\ref{pr1})
is the solution to equations (\ref{ft}) and (\ref{cTBA2})
which are equivalent, after the change of variables
$y_0=1-1/f_g(x_0)$, to equation (\ref{cTBA}) and
the relation $y_0=x_0 (1-y_0)^g$. 
\ep

An interesting feature of identity (\ref{Gsum}) is
that, although its l.h.s.~involves $g$ and $\Phi$ in
essentially different ways, its r.h.s.~depends only on 
the value of~$\nu\equiv (g+\Phi)$. Thus, for a fixed 
$y_0$, identity (\ref{Gsum}) provides a representation 
for dilogarithm $L(y_0)$ as a series with a free 
parameter. As an example, consider three special cases,
namely, $\nu = 2, 1, \frac 12$. For these values
we have, respectively, $y_0=1-\rho,\frac 12,\rho$,
where \hbox{$\rho=(\sqrt{5}-1)/2$}. 
It is known (see, e.g., \cite{Lev}) that these are the
only algebraic points on the interval $(0,1)$, where 
$\frac{6}{\pi^2} L(y_0)$ takes rational values (which 
are $\frac 25$, $\frac 12$, $\frac 35$, respectively).
Thus, keeping $g$ as a free parameter, we obtain
for the special values of $\nu$ the following identities
\begin{eqnarray}
 \label{s2a}
 && \sum_{n=1}^\infty \sin \pi g n \,
 \frac{\Gamma((1-g)n)\, \Gamma(gn)}{\pi \, n \, n!} \, 
 \rho^{(2-g)n}  +  \frac{2-g}{2} (\ln \rho)^2  =  
 \frac{\pi^2}{15}   \,, \\
  \label{s2b}
 && \sum_{n=1}^\infty \sin \pi g n \,
 \frac{\Gamma((1-g)n)\, \Gamma(gn)}{\pi \, n \, n!} \, 
  2^{(g-1)n}  +  \frac{1-g}{2} (\ln 2)^2  =  
 \frac{\pi^2}{12}   \,, \\
 \label{s2c}
 && \sum_{n=1}^\infty \sin \pi g n \,
 \frac{\Gamma((1-g)n)\, \Gamma(gn)}{\pi \, n \, n!} \, 
 \rho^{(1-2g)n}  +  (1-2g) (\ln \rho)^2  =  
 \frac{\pi^2}{10}   \,.
\end{eqnarray}
Here $0< g <1$ in (\ref{s2a})--(\ref{s2b}),
whereas the upper bound for $g$ in (\ref{s2c})
is determined from the convergence condition 
(\ref{t0}) (approximately, $g<0.88$).

\section{Gentile statistics}
Another interesting case of extensive statistics
(which appeared already in \cite{Ge} and is sometimes 
called the Gentile statistics) arises when we chose 
in (\ref{fgen}) the following generating function
\begin{equation}\label{Gf}
 F_\sG(t) = 1+t+t^2+\ldots + t^G \,,
\end{equation}
which also interpolates between fermions
($G=1$) and bosons ($G=\infty$). For this statistics,
the general formula (\ref{c2}) for the effective
central charge acquires the form \cite{By}
\begin{equation}\label{cG}
  \tilde{c}_\sG
  = \frac{6}{\pi^2} \Bigl[ L \left( x_0 \right) - \frac{1}{G+1}
    L \left( x_0^{G+1} \right) \Bigr] \,,
\end{equation}
where  $x_0$ is the positive root of equation
(\ref{cG2}) for $f(t)=F_\sG(t)$.

The maximal value of $\mu$ for which equation (\ref{xeq})
has positive root is interpreted (because $\mu =n/N$ in
formula (\ref{ent1})) as the maximal occupation number 
for a single state. It is easy to see that this number
is $\mu_{\rm max}=G$ for the Gentile statistics and
$\mu_{\rm max}=1/g$ for the Haldane--Wu statistics.
In both cases the entropy density $s(\mu)$ is a concave
function such that $s(0)=s(\mu_{\rm max})=0$. Therefore,
it is natural to compare properties of the
Gentile statistics with parameter $G$ and the Haldane--Wu
statistics with parameter~$g=1/G$. It was conjectured in
\cite{By} that the former statistics majorizes the latter.
Here we will prove the following statement.
\begin{pro}\label{P3}
Let $1<G<\infty$ and $g=1/G$. 
Then the Gentile statistics majorizes the 
Haldane--Wu statistics in the sense that 
\begin{equation}\label{maj1}
    F_\sG(t) > f_g(t)  
\end{equation}
for $t>0$.
\end{pro}

\bp To prove this assertion, it is again useful to use
the function $y(t)=1-1/f_g(t)$; equation (\ref{ft})
then acquires the form $t=y(1-y)^{-g}$.  Hence
\begin{equation}\label{Ff}
 (1-t)\, \bigl(F_{\frac1g}(t) - f_g(t)\bigr) = 
 1 - t^{1+\frac 1g} + (t-1)\, f_g(t) = 
 y \, (1-y)^{-g-1} \, \phi_g(y) \,,
\end{equation}
where $\phi_g(y) = 1 - y^{\frac 1g} - (1-y)^g$. Let us
show that $\phi_g(y)>0$ for $0<t<1$, i.e., for $0<y<y_0$, 
where $y_0$ is the positive root of equation $y_0=(1-y_0)^g$.
The inequality
\begin{equation}\label{yi}
   (1-y)^g < 1-gy \,, 
\end{equation}
that holds for $0<g<1$ and $0<y\leq 1$, leads to
the estimate $\phi_g(y)> gy - y^{\frac 1g}$.
Consequently, $\phi_g(y)>0$ for $0<y\leq\tilde{y}$, 
where $\tilde{y}$ is the positive root of equation
$g\tilde{y}={\tilde{y}}^{\frac 1g}$. 
For $y>\tilde{y}$ we find
\begin{equation}\label{phii}
  \phi_g^\prime(y) = g(1-y)^{g-1} - \frac{1}{g}y^{\frac 1g -1}
 < g(1-y)^{g-1} - 1 < - \frac{1-g}{1-y} \bigl(1-y(1+g)\bigr) \,, 
\end{equation}
where we again used inequality~(\ref{yi}). On the other
hand, for $y \leq y_0$, it follows from (\ref{yi}) that 
$y < y_0 < \frac{1}{1+g}$. Therefore $\phi_g^\prime(y) < 0$ 
on the interval $\tilde{y} < y < y_0$. And since
$\phi_g(y_0)=1 - y_0^{\frac 1g} - (1-y_0)^g=1-y_0^{\frac 1g}
 -y_0 = 0$, we conclude that $\phi_g(y)>0$ on this interval
as well. Thus, the r.h.s.~of (\ref{Ff}) is positive for 
$0<t<1$. Using that $\phi_g(y)=\phi_{\frac 1g}(1-y)$
(notice that inequality (\ref{yi}) reverses for $g>1$), 
we can analogously show that the
r.h.s.~of (\ref{Ff}) is negative for $t>1$. Finally,
for $t=1$ we have (also cf.~(\ref{fas}))
\begin{equation}\label{t1}
 f_g(1) = \frac{1}{1-y_0} < 1 + \frac{1}{g} 
 = F_{\frac 1g}(1) \,,
\end{equation}
which completes the proof. \ep

Let us remark that, as seen from the proof, $G$ in 
(\ref{maj1}) does not have to be an integer if we write 
$F_\sG(t)$ as $(1-t^{G+1})/(1-t)$. 
Actually, doing so, we can consider also the case
$0<G<1$. In this case inequality (\ref{maj1})
reverses as can be shown by a proper modification of 
the above proof. However, for a physical interpretation, 
the case of non-integer $G$ is less natural.

Proposition~\ref{P3} can be used to establish
inequalities between physical quantities related to 
the statistics in question. For example, we will prove
the following. 
\begin{pro}\label{P4} 
Let $\tilde{s}_\sG(\mu)$ and $\tilde{c}_\sG$ 
be the entropy density and the effective central charge
for the Gentile statistics, and let $s_g(\mu)$ and $c_g$ 
be the entropy density and the effective central charge
for the Haldane--Wu statistics. Then, for $1<G<\infty$ 
and $g=1/G$,  the following inequalities hold
\begin{eqnarray}
 \label{maj2}
 & \tilde{s}_\sG(\mu) > s_g(\mu) \,,  &  \\
 \label{maj3}
 & \tilde{c}_\sG > c_g \,, &
\end{eqnarray}
where $0 < \mu < G$ in (\ref{maj2}).
\end{pro}

\bp
For fixed value of $\mu$, equations (\ref{ent2}) and
(\ref{xeq}) define the entropy density as a functional
of the generating function, $s=s[f]$. Taking a small
variation of the function $f$ (which involves also
variation of $x$ via (\ref{xeq})), we obtain
\begin{equation}\label{vars}
  \delta s[f] = \delta (\ln f - \mu \ln x)
 = \frac{\delta f}{f} + \frac{f^\prime}{f} \delta x
 - \frac{\mu}{x} \delta x = \frac{\delta f}{f} \,, 
\end{equation}
where the last equality took into account 
equation~(\ref{xeq}). 

Analogously, for fixed value of $\Phi$, equations 
(\ref{c2}) and (\ref{cG2}) define the functional $c[f]$.
For a small variation of the function $f$ we find
\begin{equation}\label{varc}
 \delta \Bigl( \frac{\pi^2}{6} \, c[f] \Bigr) = 
 \frac{1}{2} \Bigl( \frac{\ln f(x_0)}{x_0} \, \delta x
 - \frac{\ln x_0}{f(x_0)} \, \delta f - 
 \ln(x_0) \frac{f^\prime (x_0)}{f(x_0)} \, \delta x \Bigr) +
 \int_0^{x_0} \frac{ dt }{t} \delta f(t) =
 \int_0^{x_0} \frac{ dt }{t} \delta f(t) \,,
\end{equation}
where we used equation (\ref{cG2}) and its
consequence, $f(x_0) \, \delta x + \Phi \, x_0 \, 
 (\delta f + f^\prime (x_0) \, \delta x) =0$.

Now consider 
$\psi_a(t)= a F_{\frac 1g}(t) + (1-a) f_g(t)$ for
$a\in [0,1]$. This function is positive for all $t$
and moreover, due to Proposition~\ref{P3},
$\delta \psi_a(t)= \delta a \,
 \bigl( F_{\frac 1g}(t) -  f_g(t) \bigr) > 0 $ if 
$\delta a> 0 $ and $t>0$. 
This, together with (\ref{vars})--(\ref{varc}), 
implies that $s[\psi_a]$ and $c[\psi_a]$ are 
monotonically growing functions of $a$ and hence 
relations (\ref{maj2})--(\ref{maj3}) follow. \ep

Relation (\ref{maj3}) gives us an inequality 
involving the Rogers dilogarithm at specific 
arguments. Let us formulate it explicitly.
\begin{pro}\label{P5} 
Let $\Phi \geq 0$ and $0\leq g \leq 1$.
Let $x_0$ and $y_0$ be, respectively, the positive
roots of equations
\begin{eqnarray}
 \label{cF}
 && \ln x_0 =  \Phi \, \ln(1-x_0) -
 \Phi \, \ln \Bigl( 1-x_0^{1+\frac 1g} \Bigr) \,,  \\
 \label{cf}
 && \ln y_0 = ( \Phi + g ) \, \ln (1-y_0) \,. 
\end{eqnarray}
Then
\begin{equation}\label{Lin}
 L \bigl( x_0 \bigr) - \frac{g}{1+g}
    L \Bigl( x_0^{1+\frac 1g} \Bigr) \geq
 L(y_0) 
\end{equation}
and the equality takes place if and only if
$g=0$ or $g=1$.
\end{pro}

\bp For $g=0^+$ we have $x^{\frac 1g}=0$. Then
equations (\ref{cF})--(\ref{cf}) yield $y_0=x_0$
and (\ref{Lin}) is obviously an equality.
 For $g=1$ equations (\ref{cF})--(\ref{cf}) yield 
 $y_0= \frac{x_0}{1+x_0}$. Then (\ref{Lin}) 
becomes an equality due to the Abel identity
$L(t^2)=2L(t) - 2 L(\frac{t}{1+t})$ that holds
for any $t$ on the interval $[0,1]$.

 For $0< g < 1$ the inequality in (\ref{Lin}) follows 
{}from relation (\ref{maj3}) in Proposition~\ref{P4},
equations (\ref{cHW})--(\ref{cTBA}), formula 
(\ref{cG}), and equation (\ref{cG2}) for
$f(t)=(1-t^{1+\frac 1g})/(1-t)$. \ep

In the simplest case, $\Phi=0$, we have
$x_0=1$ and Proposition~\ref{P5} reduces to the 
estimate
\begin{equation}\label{L1}
  L(y) < \frac{1}{1+g} \frac{\pi^2}{6} \qquad
 {\rm for} \quad  y=(1-y)^g \quad
 {\rm and} \quad 0< g < 1 \,.
\end{equation}
The case $\Phi=1$ can be interpreted as related
to the $A_2$ affine Toda model \cite{By} and
to the Calogero--Sutherland model model with
the coupling constant $\lambda=g$~\cite{BF}.

\vspace*{1.5mm}
{\em Remark}. After this manuscript had been written 
the author was informed that multivariable analogues 
of formulae (\ref{wn}) and (\ref{fmn}) were obtained
in \cite{Ig} and \cite{KNT} by means of the
multivariable Lagrange inversion theorem.

\subsection*{Acknowledgments} The author is 
grateful to Alexander von Humboldt Foundation
for support.

\noindent
\sc Institut f\"ur theoretische Physik,
 Freie Universit\"at Berlin\\
\sc Arnimallee 14, 14195 Berlin, Germany\\
\rm and \\
\sc Steklov Institute for Mathematics\\
\sc Fontanka 27, 191011 St.Petersburg, Russia

\end{document}